\documentclass[%
reprint,
superscriptaddress,
bibnotes,
amsmath,amssymb,
aps,
longbibliography
]{revtex4-1}

\usepackage{qcircuit}
\usepackage[dvipdfmx]{graphicx}
\usepackage{graphicx}
\usepackage{amsmath,amssymb,amsthm,mathrsfs,amsfonts,dsfont}
\usepackage{subfigure, epsfig}
\usepackage{braket}
\usepackage{bm}
\usepackage{enumerate}
\usepackage[colorlinks,linkcolor=blue,citecolor=blue]{hyperref}
\usepackage{newtxtext,newtxmath}
\usepackage{algpseudocode}

\newcommand{\tr}{\mathrm{Tr}}

\newcommand{\tildeH}{\widetilde{H}}
\newcommand{\tildeS}{\widetilde{S}}

\graphicspath{{./fig/}}  

\usepackage{graphicx}
\usepackage{dcolumn}
\usepackage{bm}
\usepackage{xcolor}

\begin{document}

\preprint{APS/123-QED}
\title{
Balancing error budget for fermionic $k$-RDM estimation
}


\author{Nayuta Takemori}
\email{nayuta.takemori.qiqb@osaka-u.ac.jp}
\affiliation{Center for Quantum Information and Quantum Biology, Osaka University, 1-2 Machikaneyama, Toyonaka, Osaka, 560-0043, Japan}
\affiliation{Center for Emergent Matter Science, RIKEN, Wako, Saitama 351-0198, Japan}

\author{Yusuke Teranishi}
\affiliation{Graduate School of Information Science and Technology, Osaka University, 1-5 Yamadaoka,Suita,Osaka,565-0871,Japan}

\author{Wataru Mizukami}
\affiliation{Center for Quantum Information and Quantum Biology, Osaka University, 1-2 Machikaneyama, Toyonaka, Osaka, 560-0043, Japan}
\affiliation{Graduate School of Engineering Science, Osaka University, 1-3 Machikaneyama, Toyonaka, Osaka 560-8531, Japan.}

\author{Nobuyuki Yoshioka}
\email{nyoshioka@ap.t.u-tokyo.ac.jp}
\affiliation{Department of Applied Physics, University of Tokyo, 7-3-1 Hongo, Bunkyo-ku, Tokyo 113-8656, Japan}
\affiliation{Theoretical Quantum Physics Laboratory, RIKEN Cluster for Pioneering Research (CPR), Wako-shi, Saitama 351-0198, Japan}
\affiliation{JST, PRESTO, 4-1-8 Honcho, Kawaguchi, Saitama, 332-0012, Japan}

\date{\today}

\begin{abstract}
The reduced density matrix (RDM) is crucial in quantum many-body systems for understanding physical properties, including all local physical quantity information.
This study aims to minimize various error constraints that causes challenges in higher-order RDMs estimation in quantum computing.
We identify the optimal balance between statistical and systematic errors in higher-order RDM estimation in particular when cumulant expansion is used to suppress the sample complexity. Furthermore, we show via numerical demonstration of quantum subspace methods for one and two dimensional Fermi Hubbard model that, biased yet efficient estimations better suppress hardware noise in excited state calculations.
Our work paves a path towards cost-efficient practical quantum computing that in reality is constrained by multiple aspects of errors.
\end{abstract}
\maketitle


{\it Introduction.---} 
The design, implementation, and utilization of quantum computers have attracted growing interest~\cite{shor1999polynomial, shor1995scheme,  nielsen2002quantum, lidar2013quantum}. Fields such as condensed matter physics and quantum chemistry offer prime testing grounds for understanding the principles of these quantum devices~\cite{lloyd1996universal, abrams1999quantum, aspuru2005simulated, babbush_encoding_2018, lee_evenmore_2021, Yoshioka2022}. However, current quantum computing technologies with gate fidelities typically around $10^{-3}$ pose considerable challenges even with the methodologies to combat noise without error correction~\cite{temme2017error, li2017efficient, koczor2020exponential, huggins2020virtual, yoshioka2022generalized, cai2023quantum,  endo2021hybrid, tsubouchi2023universal, takagi2023universal, quek2022exponentially}. These devices, while being a significant step forward, are modest in scale and lack the necessary qubits for achieving fault tolerance, which assumes logical error rates to below $10^{-10}$, for instance. 
While there has been rapid advancement in developing logical qubits~\cite{ofek2016extending, anderson2021realization, krinner2022realizing, sivak2023real,  sundaresan2022matching, bluvstein2023logical, google2023suppressing}, it is envisioned that the coming era would be rather early fault-tolerance whose gate operations are more accurate than the physical ones yet not precise as described above. Therefore, it is crucial to reveal how to distribute the budget of errors from various resources for practical usage of quantum computers.


One critical area in quantum computing is the study and effective handling of electron correlation effects in quantum systems, which is fundamental in condensed matter physics and quantum chemistry. These effects significantly influence the behavior of materials and molecules.
To consider electron correlation effects, the reduced density matrix (RDM) in quantum many-body systems is extremely useful in elaborating physical properties, as it contains all the local information. In particular, higher-order RDMs play a pivotal role in quantum algorithms that accurately account for electron correlation~\cite{Yanai2007}, however, the computation of higher-order RDMs, especially those above the third order, face a significant challenge due to the scalability issue. 
The number of terms in the $k$-RDM increases on the order of $O(N^{2k})$ with $N$ representing the number of qubits or the number of fermionic modes, and it has been pointed out that the sample complexity of $O(N^k)$ is unavoidable~\cite{bonet-monroig2020nearly, cotler2020quantum}.

 One of the state-of-the-art measurement scheme is the classical shadow tomography~\cite{aaronson2018shadow, Huang2020} that can estimate $k$-RDM of a multiqubit quantum state from $O(N^k/\epsilon^2)$ measurements within additive error of $\epsilon$.
 The initial flaw of inapplicability to fermionic systems was resolved by subsequent works that utilized fermionic Gaussian unitaries~\cite{Zhao2021a}, matchgates~\cite{Wan2022}, or polynomial interpolation of Pfaffians~\cite{low2022classical}, to achieve complexity of $O(N^k/\epsilon^2)$.
 While these works achieve the near-optimal complexity under projective measurements, they all deal with the asymptotic and noiseless limit such that statistical error is the only source of error to be taken into account.
 Considering the versatility of error resource, such as statistical, hardware and algorithmic errors, that are to be minimized as a total rather than individual, one must compile various error budget to enhance the utility of quantum computers.

In this paper, we investigate how to minimize the holistic effect of errors that are present during computation that utilizes higher-order RDMs. 
We first  reveal the optimal tradeoff between the statistical and systematic errors by utilizing the cumulant expansion to estimate higher order RDMs from low-order ones. 
In particular, we compute the error in estimating 3-RDMs for random state and argue that, under realistic amount of measurement budget, the cumulant-based estimation is more precise than the unbiased estimation using fermionic shadow tomography.
Furthermore, we demonstrate the biased yet measurement-efficiant estimation via excited state calculation based on quantum subspace methods for one and two-dimensional Fermi Hubbard model and show it is capable of suppressing the effect of hardware noise than unbiased optimal measurement strategy.
Our work paves a path towards 
 cost-efficient practical quantum computing that in reality is constrained by multiple aspects of errors. 

{\it Efficient estimation of fermionic RDMs}.---
First, we briefly review the fermionic shadow tomography~\cite{Zhao2021a} which is a near-optimal measurement protocol to estimate fermionic RDMs in an unbiased manner.
Let us first define a tensor element of a fermionic $k$-RDM for the quantum state $\rho$ as 
\begin{equation}
\label{eq:krdm}
    ^kD^{i_1 \cdots i_k}_{j_1 \cdots, j_k} = \frac{1}{k!}{\rm Tr}[a_{i_1}^\dag \cdots a_{i_k}^\dag a_{j_k} \cdots a_{j_1} \rho],
\end{equation}
where $a_{i}^{(\dag)}$ is the annihilation (creation) operator of $i$-th fermionic mode. Note that this imposes antisymmetric relation between the tensor elements. 
In order to estimate the elements on the quantum computer, we introduce the 
Majorana operator representation of fermionic operators as
\begin{equation}
\gamma_{2p}=a_{p}+a_{p}^{\dagger},\ \ \gamma_{2p+1}=-i(a_p-a_p^{\dagger}),
\end{equation}
which is convenient since it has the same algeraic properties as the Pauli operators.
For each $2k$-combination ${\bf \mu}=(\mu_1,\cdots,\mu_{2k}) \in \mathcal{C}_{2N, 2k}$, 2$k$-order Majorana operator can be given by
\begin{align}
\Gamma_{\mu}=(-i)^k\gamma_{\mu_1}\cdots \gamma_{\mu_{2k}}.
\end{align}



To estimate the expectation value for all $\Gamma_{\mu}$ up to $2k$-th order, we perform measurements on randomized basis that is tailored for fermionic systems. Concretely, for each measurement, we randomly choose a unitary matrix $U$ from a subset of the fermionic Gaussian unitary group $\mathrm{FGU}(N)$,
which consists of all the following unitary matrices:
\begin{align}
U(e^A)=\mathrm{exp}\left ( -\frac{1}{4}\sum^{2N-1}_{\mu,\nu=0}A_{\mu\nu}\gamma_{\mu} \gamma_{\nu} \right ),
\end{align}
where $A = -A^T \in \mathbb{R}^{2N\times 2N}$.
In particular, here we impose particle number conservation, which results in the ensemble composed from the alternating group $\mathrm{Alt(2N)}$ as
\begin{align}
\mathcal{U}_{\mathrm{FGU}}=\left\{ U(Q) \in \mathrm{FGU}(N) | Q\in \mathrm{Alt}(2N) \right\},
\end{align}
whose accompanying action can be given by $U(Q)^{\dagger}\gamma_{\mu}U(Q)=\sum_{\nu=0}^{2N-1}Q_{\mu \nu} \gamma_{\nu}.$

Upon measuring the mapped Majorana operators under the basis determined by the matrix $Q$, one obtains the computational basis $|z\rangle$ where $z \in \{0,1\}^N$. The corresponding estimator $\hat{\rho}_{Q, z}$, or the classical shadow, can then be expressed as
\begin{align}
\mathrm{tr}(\Gamma_{\mu}\hat{\rho}_{Q,z})=\lambda^{-1}_{N,k}\sum_{\nu \in C_{2N,2k}}\langle z | \Gamma_{\nu} |z\rangle\mathrm{det}[Q_{\nu,\mu}],
\end{align}
where $Q_{\nu,\mu}$ refers to the submatrix of $Q$~\cite{chapman2018classical}, $\lambda_{N,k} = \begin{pmatrix}N\\k\end{pmatrix} / \begin{pmatrix}
    2N \\ 2k
\end{pmatrix}$.
After repeated measurements, the expected value for the physical quantity $\Gamma_{\mu}$ can be given by average over the sampled $Q$ and the measurement outcomes as
\begin{equation}
    \langle \Gamma_{\mu}\rangle = \mathbb{E}_{Q,z} \left[\mathrm{tr}[\Gamma_{\mu}\hat{\rho}_{Q, z}] \right]. \label{eq:krdm_shadow}
\end{equation}


By using Eq.~\eqref{eq:krdm_shadow}, we may estimate the $k$-RDM in an unbiased way with sample complexity of $O(N^k/\epsilon^2)$.
While this is polynomial with $N$, the practical number of measurement shot scales prohibitively large for higher-order RDMs~\cite{nishio2023statistical}.
One of the workarounds considered in the context of quantum chemistry is to estimate higher-order RDMs from the lower-order ones using the celebrated cumulant expansion method  (See Supplemental Material (SM)~\cite{see_sm} for details).
This approach is based on the following decomposition of RDMs,
\begin{align}
{}^1D&={}^1\Delta \\
{}^2D&={}^2\Delta+{}^1\Delta \wedge {}^1\Delta \\
{}^3D&={}^3\Delta+3{}^2\Delta\wedge {}^1\Delta +{}^1\Delta \wedge {}^1\Delta  \wedge {}^1\Delta 
\end{align}
where $\wedge$ denotes the wedge product and ${}^k\Delta$ denotes the $k$-th order connected RDMs~\cite{MAZZIOTTI1998419}. 
The cumulant-based approximation of RDMs is performed by neglecting the higher order connected RDMs, such that we may substitute the direct estimation of them by summation over lower-order ones.
For instance, by neglecting $\Delta^3$ and solving the above relationship recursively, one can obtain the approximation of the 3-RDM as
\begin{align}
{}^3D
= 3 {}^2D \wedge {}^1D - 2 {}^1D^3.
\end{align}
This formulation significantly simplifies the process of obtaining higher-order RDMs, thereby enhancing the efficiency of quantum computational simulations. 
It has been recognized that, by leveraging the cumulant expansion, one can effectively navigate the complexities associated with high-order quantum correlations, crucial for accurate quantum system evaluations
\cite{colmenero1993approximating, nakatsuji1996direct, yasuda1997direct}.
However, the serious drawback of the cumulant expansion is that it does not have any nontrivial error bound for its performance.
In particular, it remains an open question how to determine whether or not one benefit from the use of the biased estimation.

\begin{figure}[tb]
    \begin{center}
    \includegraphics[width=0.85\linewidth]{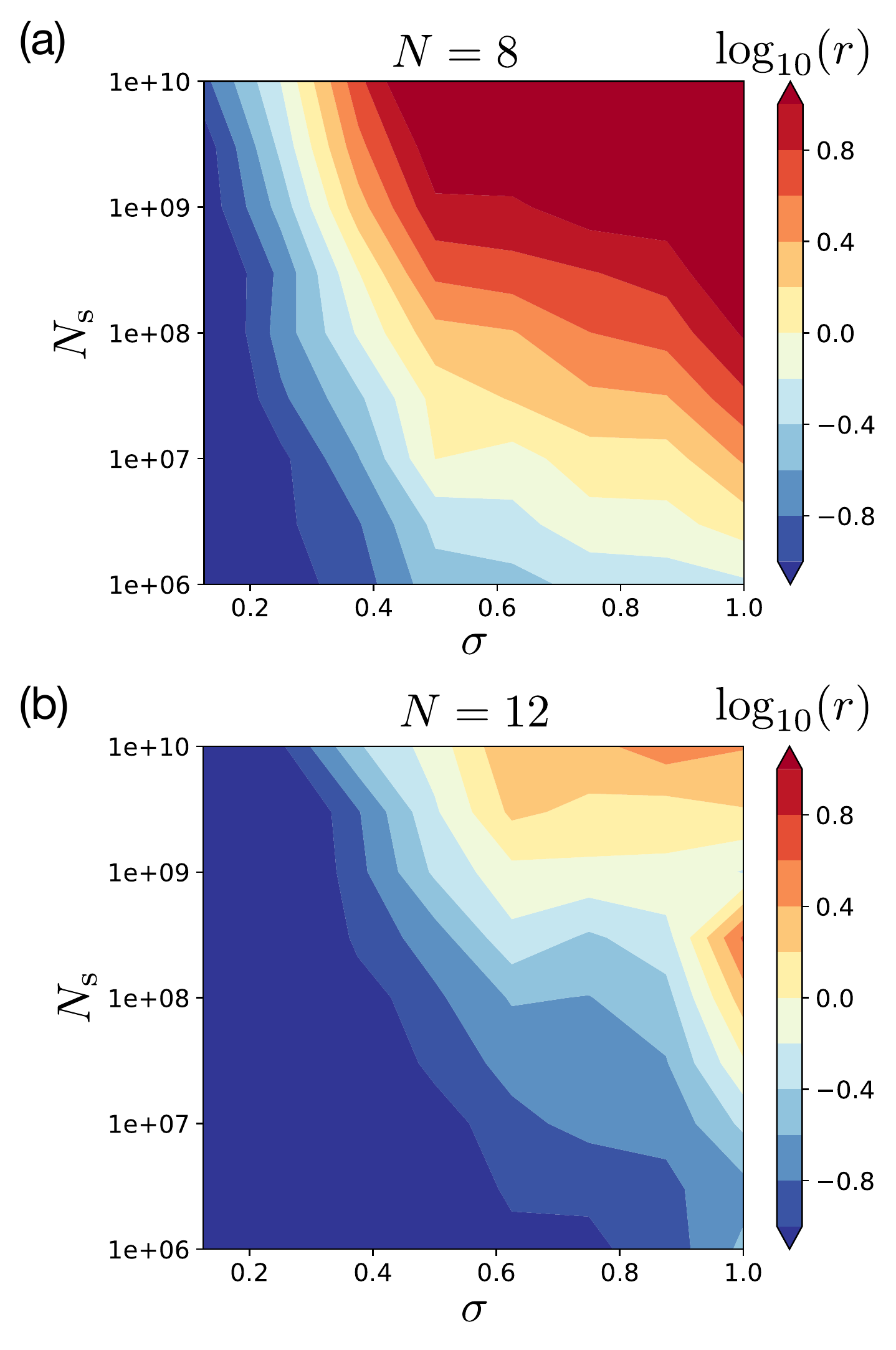}
    \caption{Comparison of 3-RDM estimation errors $\sigma$ between naive measurements using fermionic shadows and that based on cumulant expansion for random UCCSD state. The number of fermionic modes is (a) $N=8$ and (b) $N=12$ with half-filling. As the number of samples $N_s$ increases, the accuracy ratio $r$ also increases, indicating that the naive estimation is precise enough to outperform the cumulant expansion that inevitably introduces bias. Given that the cumulant-based estimation requires $O(N^2)$ samples where the naive one needs $O(N^3),$ the crosspoint $r=1$ is expected to demand even larger $N_s$ for larger system sizes.
    }\label{fig:error-uccsd}
    \end{center}
\end{figure}

{\it Error in $k$-RDM estimation.---}
First, we discuss the validity of utilizing the cumulant expansion. While the cumulant expansion partially neglects higher-order electronic correlation and hence does not yield unbiased estimation on the RDMs, we expect that there is a regime where the statistical error for the naive estimation is too large so that the bias from the cumulant expansion is less impactful.
To quantitatively search for such a regime, here we define the following estimation accuracy ratio for $k$-RDM:
\begin{equation}
    r_k = \frac{\left|^kD - {}^k\hat{D}_{\rm cum}\right|}{\left|^kD - {}^k\hat{D}\right|}, \label{eq:r_def}
\end{equation}
where $|\cdot|$ denotes the summation over absolute values of all the elements. 
We have discriminated the estimated RDM as $^k\hat{D}$ from its exact $^kD$, and introduced $^k\hat{D}_{\rm cum}$ as the estimation using the cumulant expansion.
Note that Eq.~\eqref{eq:r_def} is with respect to a single instance of a quantum state and set of measurement outcomes.
Our goal is to reveal the average-case behavior of $r$ for realistic target quantum states.

Here, we study the behavior of $r$ under estimation for randomly-parametrized Unitary Coupled-Cluster Singles and Doubles (UCCSD) ansatz using a set of parameters $\theta$ as
\begin{equation}
    |\psi_\theta\rangle = e^{T_\theta -T_\theta^\dagger} |{\rm HF}\rangle,
\end{equation}
where $|{\rm HF}\rangle = \prod_{a\in\Lambda} a_i^\dagger |0\rangle$ is a Hartree-Fock state for a subset of sites $\Lambda=\{1, ..., \lceil N/2\rceil\}$ and $T_\theta = T_\theta^{(1)} + T_\theta^{(2)}$ is a summation of single and double excitations defined as
\begin{eqnarray}
    T_\theta^{(1)} &=& \sum_{i\in{\rm unocc.}}\sum_{m\in {\rm occ.}}
    t_{im}^{(1)} a_i^\dagger a_m,\\
    T_\theta^{(2)} &=& \sum_{i,j\in {\rm unocc.}} \sum_{m,n\in{\rm occ.}} t_{ijmn}^{(2)}a_i^\dagger a_j^\dagger a_m a_n,
\end{eqnarray}
with variational parameters $t_{im}^{(1)}, t_{ijmn}^{(2)}\in \mathbb{R}$ that constitute $\theta = \{t^{(1)}_{im}\}_{im} \cup \{t^{(2)}_{ijmn}\}_{ijmn}$.
In the following, we sample each element of $\theta$ from the normal distribution $\mathcal{N}(0,\sigma)$ where $\sigma$ is the standard deviation, and consider average over instances. Furthermore, we also substitute the average over all the indices of RDM with randomly sampled indices. 
Note that we may alternatively consider the random state set to be given by Haar random states with particle number conservation, while we expect that the results  do not qualitatively change.

Figure~\ref{fig:error-uccsd} shows how the randomness of the state and the measurement budget affects the accuracy ratio $r_k$ for $k=3$. Obviously, when the number of samples $N_s$ is large, the naive estimation is precise enough so that it surpasses the accuracy of the cumulant expansion. On the other hand, when $N_s$ is not sufficiently large, it is advantageous to employ the cumulant expansion. While this is obvious when there is no electronic correlation at all at $\sigma\sim 0$, we find that, even at $\sigma \sim 0.3$ where the quantum entanglement grows extensively with the system size (see SM for numerical details~\cite{see_sm}), the similar picture holds as well. One of the most remarkable points in Fig.~\ref{fig:error-uccsd} is that, with realistic amount of measurement budget, say $N_s=10^9$, we expect for typical quantum states with $N\gtrsim 10$ that it is always beneficial to employ the cumulant expansion over the naive measurement strategy.
Given that the sample complexity scales as $O(N^2/\epsilon^2)$ for cumulant expansion while the naive strategy requires $O(N^3/\epsilon^2)$, we expect that the beneficial regime enlarges for larger system sizes unless one has access to even larger measurement resource.
It is an interesting open question to study the crosspoint for $k\geq4$.


 \begin{figure}[b]
\begin{center}
\includegraphics[width=\linewidth]{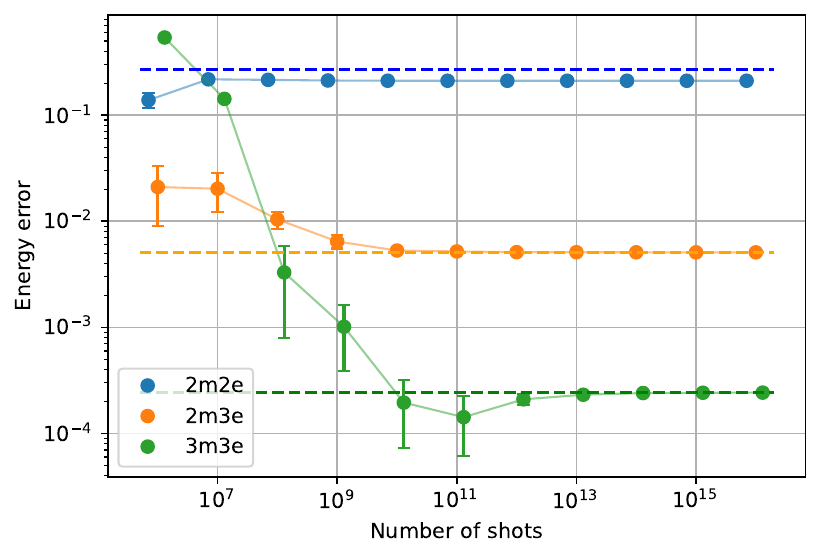}
\end{center}
\caption{\label{fig:energyerror}
Dependence of the energy error from exact solutions on the number of shots for the $N=3\times 3$ spinless Hubbard model with $U = 1$ and $\mu = 0$.
The blue, orange, green points represent the results obtained using 2-RDM estimation based on 2-RDM measurements, 3-RDM estimation using 2-RDM measurements, and 3-RDM estimation using 3-RDM measurements, correspondingly. The solid lines connecting the points illustrate the trends observed for each estimation method, while dashed lines of matching colors showcase values estimated from the exact RDM for each method.
}
\end{figure}

{\it Balancing statistical and systematic error in excited-state calculation.---}
As a useful application of the higher-order $k$-RDM estimation to quantum many-body calculation, here we consider the quantum subspace expansion (QSE)~\cite{mcclean_2017,Yoshioka2021,Yoshioka2022a, motta2023subspace}.
The QSE method is a flavour of postprocessing that virtually simulates a quantum state $\ket{\widetilde{\psi}}$ expressed as a superposition over $C_i \ket{\psi_0}$, where $C_i$ is a coupler that virtually generates quantum correlation and $\ket{\psi_0}$ is the actual quantum state to be realized on the quantum computer.
In other words, the QSE method is a variational method to optimize the following ansatz:
\begin{equation}
\label{eq:sim_wvfnc}\ket{\widetilde{\psi}} = \sum_i \alpha_i C_i \ket{\psi_0},
\end{equation}
where $\alpha_i \in \mathbb{C}$ is a coefficient that controls the accuracy of the virtual ansatz $\ket{\widetilde{\psi}}$. In particular, to solve an eigenvalue problem of a given Hamiltonian $H$, we impose the Ritz variational principle~\cite{MAGNASCO20133} to find that the optimal coefficients are given by solving the generalized eigenvalue problem as
\begin{equation}
    \tildeH X = \tildeS X E,
\end{equation}
where $\tildeH_{ij} = \braket{\psi_0 | C_i^\dagger \hat{H} C_j | \psi_0}$ and $\tildeS_{ij} = \braket{\psi_0 | C_i^\dagger C_j | \psi_0}$ are the modified Hamiltonian on the subspace ${\rm Span}\{C_i \ket{\psi_0}\}$ and the overlap matrix of bases for the truncated subspace, respectively.
Here we have denoted $X$ as the collection of the eigenvectors, and $E$ as a diagonal matrix that gives the eigenvalues.
Note that the matrix elements $\tildeH_{ij}, \tildeS_{ij}$ are measured on the quantum computers and hence involves statistical errors due to projective measurements, systematic errors due to the hardware imperfection \textit{etc}. 
Since the output from the QSE method can be easily deteriorated by such noise, it is crucial to perform appropriate normalization on the matrix elements~\cite{epperly2021theory} 
and quantum error mitigation to handle the unwanted effects. 
By implementing the $k$-RDM estimation in QSE, as formally presented in $\tildeH_{ij}, \tildeS_{ij}$, one can obtain quantum subspace expansion based on fermionic shadow tomography.


\begin{figure*}[tb]
\begin{center}
\includegraphics[width=\linewidth]{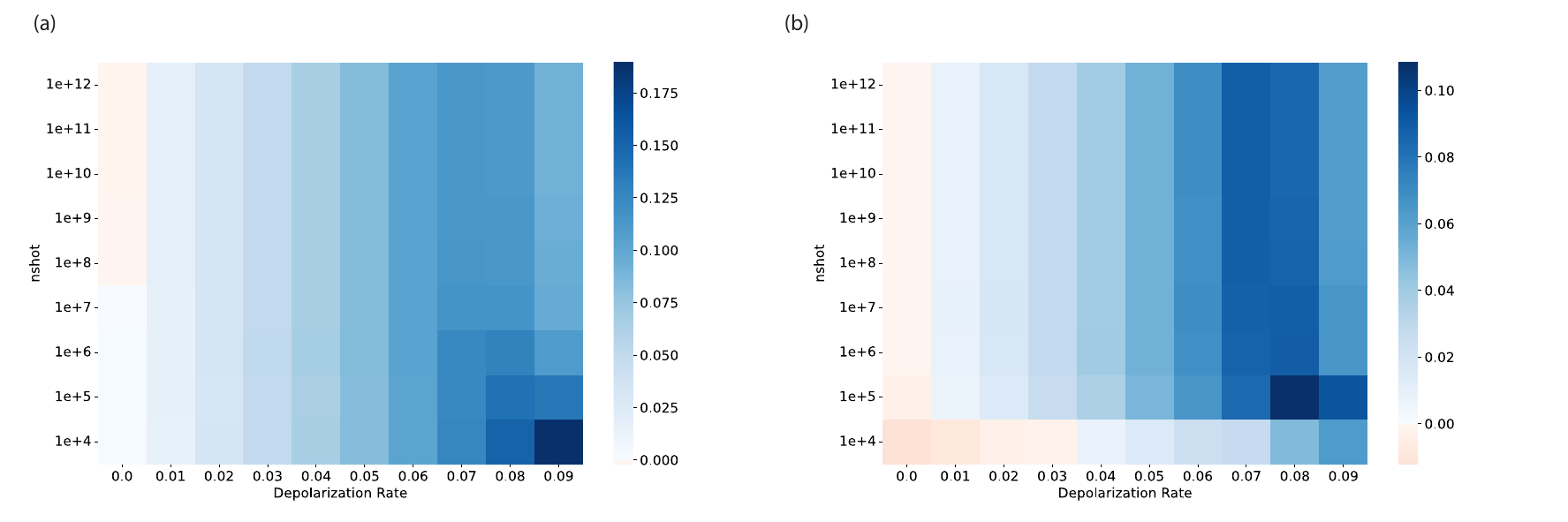}
\end{center}
\caption{\label{fig:heatmap}
Heatmap of energy error differences $E_{{\rm naive}}^{\rm QSE} - E_{\rm{cum}}^{\rm{QSE}}$ for the spinless Hubbard model. Each panel represents (a) $1d$  model with $N=9$ fermionic modes and (b) $2d$ model with $N=3 \times 3$ modes, with $U = 1$ and $\mu = 0$ for both cases. 
}
\end{figure*}

We apply our methods to a simplified model, specifically the $1d$ and $2d$ spinless Hubbard model~\cite{Prosen1999,Bera2017}.
 Recent studies have revealed that this model exhibits a range of intriguing phenomena, including dispersionless many-body bounded state~\cite{Agarwala2017}, many-body localization~\cite{deTomasi2019} and distinctive transport and dynamical properties~\cite{Bera2017}. These insights provide a deeper understanding of the model's complex behaviors and its implications in condensed matter physics. The Hamiltonian for this model is defined as:
\begin{align*}
{\cal H}=-t\sum_{\langle i,j \rangle}(a^{\dagger}_{i}a_{j}+h.c.)-\mu\sum_{i}n_{i} 
+U\sum_{\langle i,j \rangle}n_{i}n_{j}.
\end{align*}
Here, $\langle i,j \rangle$ denotes the nearest-neighbor site, $t$ is the transfer integral between sites, $\mu$ is the chemical potential $U$ is the Coulomb interaction and $n_{i}=a^{\dagger}_{i}a_{i}$.
We suppose a finite electron-transfer integral $t$ only between the nearest neighbor sites and set it as the unit of energy.
In the following, we fix $U=1$ and $\mu=0$, focusing on system sizes of $N=9$ for the one-dimensional ($1d$) model and $3 \times 3$ for the two-dimensional ($2d$) model. Here, we choose $|\psi_0 \rangle$ to the exact solution on the $n$ particle sector where $n$ represents the number of particles for the ground state, and $C_i=a_i$ a coupler, which leads us to focus on the $n -1$ particle sector as a virtual state.




Figure~\ref{fig:energyerror} shows the tradeoff between the statistical and systematic error in excited-state energy calculation for the $2d$ case. The energy error is compared to its exact solution of the $n-1$ particle sector. 
We compare the three different evaluations: 2-RDM estimation based on 2-RDM measurements, \textit{i.e.}, without using cumulant expansion, 3-RDM estimation using 2-RDM measurements, and 3-RDM estimation using 3-RDM measurements. The solid lines are a guide to the eyes, while dashed lines of matching colors showcase values estimated from the exact RDM for each method. Note that the asymptotic values from all estimation methods do not match to exact zero since the couplers $\{a_i\}$ take only single-particle excitations and thus the variational form of Eq.~\eqref{eq:sim_wvfnc} is not expressive enough to encode the exact excited state. As the number of shots is increased, the values approaches that are obtained from a rigorous RDM evaluation. As can be expected from the results for the $k$-RDM estimations as in Fig.~\ref{fig:error-uccsd}, in the low-shot region, we find regions where the error is smaller when 3-RDM is evaluated by cumulant expansion rather than direct measurements. 
Therefore, we verify that the tradeoff relation holds also for many-body calculations that process the $k$-RDM estimation results in a nonlinear manner.

{\it Balancing all together in excited-state calculation.---}
Next, we consider the effect of global depolarizing noise, which is a common form of hardware noise that arise under sufficiently deep quantum circuits~\cite{arute2019quantum, dalzell2021random}.
As shown in Fig.~\ref{fig:heatmap}(a), in the absence of hardware noise (\textit{i.e.}, a rate of 0), the energy error between two estimation methods—the naive estimation and the estimation using the cumulant expansion obtained by QSE—tends to decrease with an increasing number of shots, eventually converging to the exact value of the RDM. Conversely, in the presence of hardware noise, the range of shots where the cumulant expansion-based estimation results in smaller errors becomes more extensive. This observation highlights the significant impact of hardware noise on the accuracy of quantum measurements.
Interestingly, the presence of hardware noise emphasize the advantage of utilizing the cumulant expansion in one-dimensional systems, as shown in Fig.~\ref{fig:heatmap}(b). 
The results underscore the robustness of the cumulant expansion approach in both $1d$ and $2d$ case, even in the presence of hardware noise.

{\it Discussions and outlooks.---}
 In this paper, we have studied the tradeoff between error resources in quantum computation that involves higher-order  fermionic $k$-RDMs. We have found that, in contrary to the ordinary setups where unbiased estimation is the sole option, it is beneficial in practical situations to utilize the cumulant expansion with systematic error instead of enduring the high-polynomial scaling of measurement shots.
 By further investigating the many-body calculations for excited states of spinless Hubbard model using a quantum subspace method, we find not only that the above findings hold as well, but also numerically show that the gain from cumulant expansion is stable under hardware noise.
 

We envision three future directions.
First, the extensive study for higher order RDMs is nontrivial yet crucial. Since there are numerous explorations regarding $k\geq 4$-RDMs such as contracted form of Schr\"{o}dinger equation, it is essential to evaluate the bias tradeoff relations in further practical calculations.
Second, it is highly nontrivial to derive universal tradeoff relation between the error resources studied in this work.
While there has been some attempt to investigate the tradeoff relation between hardware noise and algorithmic bias~\cite{takagi2023universal}, there is no guiding principle how to distribute the quantum resource to maximize the accuracy of the algorithm.
Third, while the current work has focused on estimations based on projective measurements, it would be highly nontrivial to extend the framework to measurements that achieve Heisenberg limit, in which the commutativity of the measured observables do not play a significant role as in the projective case. The nearly optimal measurement schemes~\cite{huggins2022nearly, van2023quantum} are to be analyzed in this direction.




{\it Acknowledgements.---}N.T., Y.T., and W.M. were supported by JST COI-NEXT program Grant No. JPMJPF2014.
N.T. wishes to thank JSPS KAKENHI Grant Nos. JP19H05817, JP19H05820, JST PRESTO Grant No. JPMJPR23F6 and the support from Yamada Science Foundation.
W.M. wishes to thank JSPS KAKENHI Grant Nos. JP23H03819 and JP21K18933,
and MEXT Quantum Leap Flagship Program (MEXT Q-LEAP) Grant Number JPMXS0120319794. 
N.Y. wishes to thank 
JST PRESTO No. JPMJPR2119, 
JST Grant Number JPMJPF2221, 
JST ERATO Grant Number  JPMJER2302, 
JST CREST Grant Number JPMJCR23I4,  
and the support from IBM Quantum.

{\it Note added.---} Upon preparation of our manuscript, we became aware of a concurrent work that use cumulant expansion combined with the fermionic shadow tomography~\cite{avdic2023fewer}.



\let\oldaddcontentsline\addcontentsline
{\renewcommand{\addcontentsline}[3]{}}
%


\clearpage

\let\addcontentsline\oldaddcontentsline
\onecolumngrid

\begin{center}
	\Large
	\textbf{Supplementary Materials for: Balancing error budget for fermionic $k$-RDM estimation}
\end{center}

\setcounter{section}{0}
\setcounter{equation}{0}
\setcounter{figure}{0}
\setcounter{table}{0}
\renewcommand{\thesection}{S\arabic{section}}
\renewcommand{\theequation}{S\arabic{equation}}
\renewcommand{\thefigure}{S\arabic{figure}}
\renewcommand{\thetable}{S\arabic{table}}

\addtocontents{toc}{\protect\setcounter{tocdepth}{0}}
\section{Cumulant expansion}
Here we review the cumulant expansion of the fermionic quantum marginals, or the reduced density matrix.
In general, the marginal of multivariate probability distribution  takes the partial trace on the subset of variables that one is interested in. For instance, given $N$ discrete variables $\{{\bf x}\}~({\bf x} = (x_1, ..., x_N))$ with homogeneous local Hilbert space dimension of $d$ that underlies a probability distribution $P({\bf x})$, the marginal distibution on the set of $k$ variables $\{x_{i_1}, ..., x_{i_k}\}$ is given as
\begin{equation}
    ^kP(x_{i_1}, ..., x_{i_k}) := \sum_{x_j:j \notin \{i_1, ..., i_k\}} P({\bf x}).
\end{equation}
This definition can be directly extended to define marginal of density matrix. Namely, under orthogonal basis spanned by $N$ local discrete variables $\{\ket{{\bf x}}\}_{\bf x}$, the marginal of the density matrix $\rho = \sum_{{\bf x}, {\bf y}} \rho_{{\bf x}, {\bf y}} \ket{{\bf x}}\bra{\bf y}$ can be given as
\begin{eqnarray}
^k\rho &=& {\rm Tr}_{\Lambda\backslash \{i_1, ..., i_k\}}[\rho],\\
    ^k\rho_{x_{i_1}, ..., x_{i_k}}^{y_{j_1}, ..., y_{j_k}} &=& \sum_{x_j: j\notin \{i_1, ..., i_k\}} \rho_{{\bf x}, {\bf y}},
\end{eqnarray}
where we denote the set of the entire sites as $\Lambda = \{i\}_{i=1}^N$.
Note that the matrix expression of the $k$-RDM no longer requires exponentially large memory of $O(d^{2N})$, but is compressed into $O(d^{2k})$.

The elements of $k$-RDM is in general different from each other, and hence the computational cost becomes prohibitively large in terms of practical sense, although it is polynomial in $N$.
This has motivated the technique of reconstructing higher order RDMs from lower order ones~\cite{mazziotti_textbook_2007, nakatsuji_yasuda_1996, yasuda_nakatsuji_1997, mazziotti_1998}.
For the sake of instructive purpose, here we provide a brief review on the general reconstruction method for $k$-RDMs proposed by Mazziotti~\cite{mazziotti_1998}.

First notice that we may define a generating functional of the RDMs as
\begin{eqnarray}
    G[J] = \tr\left[\exp(\sum_k J_k a_k^\dag) \exp(\sum_k J_l^* a_k) \rho\right],
\end{eqnarray}
where $J=(...J_k...)$ is a set of Grassman variables that are Schwinger probe to investigate the behavior of the quantum state, $J^*_k$ denotes the conjugate of the variable $J_k$, and $a_k^{(\dag)}$ is the annihilation (creation) operator of the $k$-th fermionic mode.
It can be easily shown that, by taking the partial derivative with respect to the Schwinger probes and then taking limit that all variables are set to zero, we obtain RDMs as the ``moment" of the generating functional as
\begin{eqnarray}
    \left.\frac{\partial^k G}{\partial J_{i_k} \cdots \partial J_{i_1} \partial J^*_{j_1} \cdots \partial J^*_{j_k}}\right|_{J,J^*\rightarrow 0} &=& \tr \left[a^\dag_{i_1} \cdots a^\dag _{i_k} a_{j_k} \cdots a_{i_1} \rho\right] \\
    &=& ^kD_{i_1\cdots i_k}^{j_1\cdots j_k}.
\end{eqnarray}
Next, we introduce the logarithmic of the generating functional as $G[J] = \exp(W[J])$. Following analogical discussion provided by Kubo~\cite{kubo1962generalized}, the partial derivatives of $W$ can be identified with the connected part of RDMs, which describes the higher order correlation that cannot be captured as simple product of lower order ones.
Concretely, the cumulant of the generating functional is defined as
\begin{eqnarray}
    ^k \Delta_{i_1\cdots i_k}^{j_1\cdots j_k} := \left.\frac{\partial^k W}{\partial J_{i_k} \cdots \partial J_{i_1} \partial J^*_{j_1} \cdots \partial J^*_{j_k}}\right|_{J,J^*\rightarrow 0}.
\end{eqnarray}
By examining the exponential relationship, Ref.~\cite{mazziotti_1998} showed that formulas between the moments $D$ and cumulants $\Delta$ can be derived for any order of $k$. For instance, this results in
\begin{eqnarray}
{}^1D_{i_1}^{j_1} &=&{}^1\Delta_{i_1}^{j_1}, \label{eq:cumulant_1rdm}\\
{}^2D_{i_1i_2}^{j_1j_2} &=&{}^2\Delta_{i_1i_2}^{j_1j_2} +{}^1\Delta_{i_1}^{j_1} \wedge{}^1\Delta_{i_2}^{j_2}\label{eq:cumulant_2rdm}, 
\end{eqnarray}
where we have introduced the wedge product $^1\Delta_{i_1}^{j_1}\wedge {}^1\Delta_{i_2}^{j_2} = \frac{1}{2}(^1\Delta_{i_1}^{j_1}{}^1\Delta_{i_2}^{j_2} - {}^1\Delta_{i_1}^{j_2}{}^1\Delta_{i_2}^{j_1})$ while the definition for the general case is provided below.
Observe that these equation indeed coincides with the qualitative picture that $k$-th order cumulant ${}^k\Delta$ indeed describes the ``connected" part of the $k$-RDM; the first term of Eq.~\eqref{eq:cumulant_2rdm} in the 2-RDM cannot be captured from product of 1-RDMs.
For $k=3, 4$, we further obtain the following expression:
\begin{eqnarray}
    {}^3D &=& {}^3 \Delta + 3~{}^2\Delta\wedge{}^1\Delta +{}^1 \Delta^3,\\
    {}^4 D &=& {}^4 \Delta + 4~{}^3 \Delta \wedge{}^1\Delta + 6~ {}^2\Delta\wedge{}^1\Delta^2 + 3~{}^2\Delta^2 + {}^1 \Delta^4,
\end{eqnarray}
where we have abbreviated the subscripts and superscripts for the sake of readability and used the notation of $^l \Delta^m = \underbrace{^l \Delta \wedge \cdots \wedge {}^l \Delta}_{m}$, where the wedge product between $p$-RDM and $q$-RDM in general is now defined as
\begin{eqnarray}
    (a \wedge b)_{i_1\cdots i_{p+q}}^{j_1\cdots j_{p+q}} = \frac{1}{((p+q)!)^2} \sum_{\pi, \sigma} {\rm sgn}(\pi) {\rm sgn}(\sigma) a_{\pi(i_1) \cdots \pi(i_p)}^{\sigma(j_1)\cdots \sigma(j_p)} b_{\pi(i_{p+1})\cdots \pi(i_{p+q})}^{\sigma(j_{p+1})\cdots \sigma(j_{p+q})},
\end{eqnarray}
where $\pi$ and $\sigma$ denotes a permutation on the indices with ${\rm sgn}$ indicating its sign.

The reconstruction procedure is done by simply regarding the higher connected part to be zero. For instance, if only the 1-RDM is available, the reconstruction process would neglect all the many-body correlation as $^k D \sim {}^1\Delta^k$, which corresponds to approximating the unknown quantum state as a single Slater determinant.
If one has access to the 1 and also 2-RDMs, we may estimate the 3 and 4-RDMs as 
\begin{eqnarray}
    ^3D &\sim& 3~{}^2\Delta \wedge {}^1\Delta + {}^1\Delta^3, \\
    ^4D &\sim& 6~{}^2 \Delta \wedge {}^1 \Delta^2 + 3~{}^2\Delta^2 + {}^1 \Delta^4.
\end{eqnarray}

\section{Entanglement structure in random states}
Here, we provide the numerical results on the calculation of the entanglement entropy for random states defined in the main text.

\begin{figure}[tb]
\begin{center}
\includegraphics[width=0.98\linewidth]{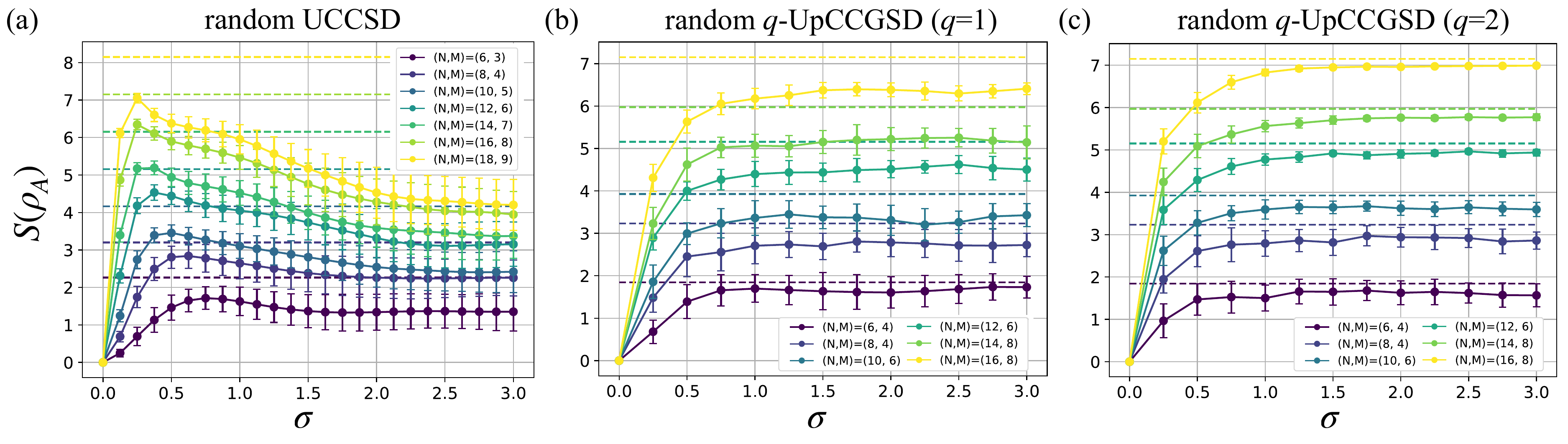}
\end{center}
\caption{\label{fig:random_uccsd_entanglement}
Half-chain entanglement entropy of (a) random UCCSD ansatz, (b, c) random $q$-UpCCGSD ($q=1,2$) ansatz averaged over 100 instances. The colored dashed lines denote the average of entanglement entropy for $N$-qubit Haar random state within the particle number sector of $M$.
}
\end{figure}

Fig.~\ref{fig:random_uccsd_entanglement} shows the half-chain entanglement entropy $S(\rho_A) = -\sum_i \lambda_i \log_2\lambda_i$ where $\{\lambda_i\}_i$ is the set of eigenvalues of the reduced density matrix $\rho_A = {\rm Tr}_{\bar{A}}[\rho]$ where $A$ denotes the half left (or right) sites of the system. Here, we consider the half-filling situation where $|{\rm HF}\rangle = \prod_{i \in A}a_i^\dag |0\rangle$ for the UCCSD ansatz defined in the main text.
By increasing the width $\sigma$ of the random parameters, we can see that the entanglement is introduced in the system which scales linearly with the system size, i.e., exhibits the volume law of entanglement.
However, large randomness is destructive in terms of the entanglement for UCCSD; the entanglement entropy peaks out above some value of $\sigma$. Within calculation up to $N=18$ fermionic modes, we observe that the entanglement is not completely degraded and saturates at some value, which also seem to scale linearly with the system size.

In contrast, as shown in Fig.~\ref{fig:random_uccsd_entanglement}(b) and (c), the destructiveness of entanglement is not observed in $q$-Unitary Pair Coupled-Cluster Generalized Singles and Doubles ($q$-UpCCGSD) ansatz. The UpCCGSD ansatz assumes that the fermionic modes have spacial and spin degrees of freedom, and is defined in the Trotterized form as 
\begin{eqnarray}
    |\psi_{\theta}\rangle = \prod_{l=1}^q \left(
    \prod_{r, s}\exp(t_{r,s}^{(1)}c_r^\dag c_s - {\rm h.c.})
    \prod_{i,j}\exp(t_{i_\alpha i_\beta j_\alpha j_\beta}^{(2)} c_{i_\alpha}^\dag c_{i_\beta}^\dag c_{j_\alpha} c_{j_\beta} - {\rm h.c.})
    \right) |{\rm HF}\rangle.
\end{eqnarray}
Here, $r,s$ run over the entire indices where $i,j$ and $\alpha, \beta$ run over the spacial and spin indices, respectively. 
While the entanglement structure exhibits qualitatively different behaviour from the UCCSD ansatz, we expect that the tradeoff relation between the statistical and algorithmic error by cumulant expansion is also present in $q$-UpCCGSD ansatz as well. This is because the tradeoff is due to the scaling difference in naive and cumulant-based estimation; as long as the quantum state captures electronic correlation, we expect that the phenomenon to be universal.

\end{document}